\documentstyle[prl,amsfonts,twocolumn,aps,epsfig]{revtex}
\begin{document}

\title{Exact diagonalization Studies of Two-dimensional Frustrated
Antiferromagnet Models}

\author{C. Lhuillier, P. Sindzingre, J.-B. Fouet}
\address{Laboratoire de Physique Th\'eorique des Liquides-UMR 7600
of CNRS,  Universit\'e  Pierre  et  Marie  Curie,  case 121,   4 place
Jussieu, 75252 Paris Cedex, France\\
E-mail:lhuilier@lptl.jussieu.fr\\ (\today)\\}
 \maketitle
\bibliographystyle{prsty}
\begin{abstract}
We describe the four kinds of behavior found in two-dimensional isotropic 
quantum antiferromagnets. 
Two of them display long range order at $T=0$: the N\'eel state and the
Valence Bond Crystal. The last two are Spin-Liquids. Properties of these
different states are shortly described and open questions are underlined.
\end{abstract}
\section{INTRODUCTION}
Exact diagonalizations are restricted to small spin systems (up to about $N=36$
spins with the present day computers). Some problems as the study of critical
regimes are out of reach of such an approach, on the contrary 
we think that this approach
might help in answering other subtle questions as for example symmetry breaking,
order out of disorder, qualitative nature of the ground-state and of the first
excitations of two-dimensional (2d) spin-1/2 magnets.

We have used it extensively to study the competition of long range order (LRO)
versus "quantum disorder" at $T=0$. Up to now, by looking
to the spectra of model spin systems, we have been able to characterize four
kinds of behavior: N\'eel LRO (collinear or not), dimer LRO 
and two different Spin-Liquids. In this paper we will give a
brief picture of these four generic cases of 2d magnetism, underline 
the more efficient numerical criteria to determine the nature
of a given unknown system and try to extract from the present numerical
knowledge what are the conditions for the appearance of the four different
phases.
The key point of our approach is the analysis of the symmetries and finite
size scaling of the low lying eigen-levels of a given hamiltonian on
small samples  with periodic (or eventually twisted) boundary conditions.
The evolution of the spectrum with the size and shapes of the sample has to
be thoroughly understood before conclusion: it may be a lengthy and sometimes
subtle process but it is usually quite valuable. Some examples will be
quoted in the following  sections.

\section{N\'eel Long range Order}

There is no mathematical proof that N\'eel long range order exists
at $T=0$ in 2d spin-1/2 systems. But there is a very
large consensus that it should be the case for the Heisenberg
nearest neighbor hamiltonian on bipartite lattices \cite{m91}. Exact
diagonalizations and/or Monte-Carlo calculations were performed for the 
Heisenberg model on the square lattice\cite{sz92}, 
on the triangular lattice\cite{blp92,cts99} and on the
honeycomb lattice\cite{rry89}. On each of these lattices the  Heisenberg
model displays long range order with a sublattice magnetization
reduced to   $60\%$ of the classical value 
on the square lattice ($\sim 40\%$ on the triangular lattice\cite{blp92,cts99},
$44\%$ on the honeycomb lattice\cite{rry89,fsl00}).
These values of the order parameter might
be in error by a few  percent: in these antiferromagnets with
linear Goldstone modes, the scaling law for the magnetization of samples
of linear size $L\sim N^{1/d}$ is an expansion in $1/L$ 
(i.e. $1/N^{1/2}$ in 2d)\cite{nz89,f89}. 
For the sizes encountered in exact diagonalizations, the asymptotic $1/N^{1/2}$
law is never reached \cite{bllp94,lblp95a} and the extrapolation to
the thermodynamic limit remains uncertain: nevertheless the
agreement between the scaling behaviors of diagonalization
results and spin wave results gives a strong support to the above-mentioned 
conclusions. 

A stronger confirmation of N\'eel LRO is obtained by the
scaling of the spectrum itself, which is an illuminating
illustration of the {\it mechanism of symmetry breaking}. 
In order to understand in a simple way how a continuous symmetry breaking 
manifests itself on finite size spectra, let us consider a
piece of solid. Our experience of the macroscopic solid is the
following: it breaks the translation symmetry and its center of
mass (c.o.m.) is localized in space. On the other hand, 
we can sort its degrees of freedom in two parts: first the
c.o.m. coordinates and the internal variables. The hamiltonian of
such a problem breaks in two commuting parts: the c.o.m. kinetic energy
and the hamiltonian of the internal degrees of freedom ( in a solid,  in a first
approximation this second part is a phonon hamiltonian). For a
finite size sample of N atoms, with mass m, this reads:
\begin{equation}
H_{tot} = \frac {{\bf P}^2}{2 m N} +h_{phonons}.
\end{equation}
The total hamiltonian does not break translation symmetry, the
total momentum ${\bf P}$ is a constant of the motion and its eigenvalues
good quantum numbers. Due to the commutation property of the
sub-parts of the hamiltonian, the eigenstates are direct products of 
eigenstates of the c.o.m. kinetic energy  and phonons
eigenstates. From these very simple observations it follows that
the spectrum of Eq. 1 as a function of the eigenvalues
of the square of the momentum has the appearance described in
Fig. 1. A straight line is drawn across the eigen-energies of the
c.o.m. hamiltonian with $0$ phonon (respectively $ 1$ phonon...).
Let us first concentrate on the vacuum of phonons: the
associated eigen-levels collapse to the absolute ground-state as
$1/N$ when N goes to infinity. Let us now consider the eigenlevels
immediately above this family: they correspond to the softest
phonon excitation and are distant from the vacuum family by the
energy of this phonon, that is an energy ${\cal O}(1/L)$
\footnote{This energy has to be measured along the arrow in Fig 1:  
this takes care of the momentum of the phonon.}.
When the size goes to infinity, {\it if the dimension of the solid is
equal or larger than two}, the levels forming the vacuum of phonons
collapse to the absolute ground-state faster than the
softest phonon! This allows to form a wave packet out of these
levels and localize the c.o.m. with a cost in energy
lower than the softest phonon: the solid phase with its specific
macroscopic properties (symmetry breaking and stiffness) can exist in 2d 
(at $T=0$), whereas it cannot exist quantum mechanically in 1d. 

Let us now turn to the N\'eel ordered magnet. For simplicity we
will first consider a collinear magnet: the macroscopic state of
the magnet at $T=0$ is defined by the Euler angles of the order
parameter. They are the equivalent of the c.o.m.
coordinates in the solid problem. The conjugate variables of these
Euler angles are the components of the total spin and their free
precession is described by the Hamiltonian:
\begin{equation}
H_{coll. var.}= \frac {{\bf S}^2}{2 \chi N}
\end{equation}
This is the equation of motion of a rigid rotor. The moment of
inertia of this rotor is written here in terms of the intensive
homogeneous spin susceptibility $\chi$, $N$ is the number of spins.
If the system has LRO and rigidity,
the motion of the internal variables is governed by an 
hamiltonian of (antiferromagnetic ) magnons and the total
hamiltonian reads:
\begin{equation}
H_{tot} = \frac {{\bf S}^2}{2 \chi N} +h_{magnons}.
\end{equation}
When $H_{tot}$ is $SU(2)$ invariant, ${\bf S}^2$ is a constant of
motion and its eigenvalues $S(S+1)$ are good quantum numbers. It 
is thus interesting to display the spectrum of this ordered antiferromagnet
as a function of $S(S+1)$. Its generic appearance is shown in Fig. 2.
One recognizes the
(Pisa-)tower of states associated with the vacuum of magnons $\left|0\right>$
(these states are called QDJS for Quasi Degenerate Joint States in refs.
\cite{blp92,bllp94}), and the tower of states associated with the softest magnon
$\left|1\right>$. The semi-classical N\'eel ground-state with an order
parameter pointing in the $( \theta, \phi)$ direction is a wave
packet of eigen-levels
with different total spin belonging to the QDJS $\left|0\right>$.
Mixing different S values up to $\sqrt N$ allows the localization
of the order parameter. If the levels of the $\left|0\right>$ family collapse
to the ground-sate faster than the softest magnon, the classical
picture survives to quantum fluctuations and the system
exhibits LRO: as for the solid, that might be the case
at $T=0$ for 2d magnets. For $T \neq 0$,  thermal
fluctuations destroy the order (Mermin-Wagner theorem). Notice
that the entropy of the semi-classical ground-state scales as
$Ln(N^2)$, and the entropy per spin of the ordered ground-state
is zero in the thermodynamic limit. 

The levels associated to the vacuum of magnons have special
properties with respect to the lattice symmetry group: these
properties reveal the space symmetry breaking of the N\'eel
state. Let us take as an example the collinear order on the
honeycomb lattice (Fig. 3): it does not break the translational
symmetry of the Bravais lattice (thus only the ${\bf k =0}$
momentum appears in the QDJS). But it breaks both the inversion
symmetry with respect to the center of an hexagon and the
reflexion with respect to an axis going through nearest neighbor
hexagon centers,  and it is invariant in the product of these two
symmetries.
Thus both the even and odd irreducible representations 
of these two symmetry  groups appear in the QDJS (symbols
$R_{\pi}=\pm 1$, $\sigma =\pm 1$), but they are always associated
so that 
\begin{equation}
R_{\pi} . \sigma = +1
\end{equation}
which insures the invariance in the product of the two symmetries, and
\begin{equation}
\sigma . (-1)^S =+1
\end{equation}
which insures the invariance of the semi-classical ground-state
in the product of a reflexion times a spin-flip.
This is the simplest situation for an antiferromagnet.
On the square lattice (which is equally bipartite) N\'eel order is
collinear and Eq. 3 remains valid, but as the translation symmetry
is broken, both the ${\bf k}= (0,0)$ and  ${\bf k}= (\pi,\pi)$
vectors appear among the QDJS.

A more complex order parameter implies a richer structure of the QDJS:
in the case of the Heisenberg model on the triangular lattice 
the order parameter is no longer a vector but a trihedron 
(or equivalently a rotation matrix).
Eq.3 is then more complex, involving both the in-plane and 
out of plane susceptibilities, and there is an internal generation of an 
extra quantum number (the projection of the total spin on the helicity
of the magnet). The free precession hamiltonian is that of a symmetric top
\cite{blp92,bllp94,adm93}
and ${\cal O}(N^3)$ states collapse to the ground-state 
in the thermodynamic limit.
(Quantum mechanically the QDJS associated to p-sublattice  N\'eel order
are adiabatically connected to
the states obtained by addition of p spins of size $\frac{N}{2p}$.)
A thorough analysis of the symmetry breaking associated to N\'eel order
on the triangular lattice is done in \cite{bllp94}.

Unexpectedly the mechanism of "order by disorder" can be detected 
by looking at the finite size scaling of relatively small samples \cite{lblp95}.
We investigated this mechanism  in the $J_1-J_2$ model on the triangular 
lattice. For $1/8<J_2/J_1<1$ the classical ground-state exhibits a large
degeneracy associated to  different kinds of 4-sublattice N\'eel order.
For small sizes ($N=16$) the quantum fluctuations are weak and
the vacuum of magnons involve all the levels associated with addition
of four momenta. In the $N=28$ sample quantum fluctuations are already
strong enough to
lift the degeneracy of this family: the levels characteristic of the
2-sublattice  N\'eel order remain the lowest ones in the
spectrum, the other ones  drift towards the continuum of
excitations. It is a nice illustration of the fact that long wave length quantum
fluctuations tend to restore the symmetries of the original
hamiltonian. The
quantum ground-state is more symmetric than the classical one.

Order parameters, spin susceptibilities \cite{bllp94}, 
stiffnesses\cite{lblp95a}
can be extracted from the numerical data: they are interesting
pieces of data to be compared to the spin waves calculations at
finite sizes, but as it has been discussed in the beginning of
this section, the leading term of their finite size scaling is
${\cal O}(N^{-\frac {1}{2}})$ and therefore it is more significant to look to
the tower of QDJS which collapse as ${\cal O}(N^{-1})$. It is also
simpler to look at the symmetries of the low lying levels and
deduce the symmetry of the order parameter than to determine it by
direct computations. As a consequence when looking at the
disappearance of N\'eel LRO due to extra frustration
or quantum fluctuations it is preferable to pinpoint the
range of parameters where the tower of states is destroyed than
search for the vanishing of the order
parameter\cite{lblps97,lmsl00}.

\section{ Dimer Long Range Order and Valence Bond Crystals}

 The second class of long-range ordered systems does not imply
$SU(2)$ symmetry breaking but only space symmetry breaking. We
will call these systems   Valence Bond Crystals (VBC) 
\footnote{The
name Valence Bond Solid is already "registered" by Affleck, Kennedy
Lieb and Tasaki \cite{aklt87,aklt88} for very specific systems
where the individual spins obey the relation $2S=z$ (with z  the
coordination number of the lattice). In such circumstances there
exists a whole class of Hamiltonians  which have  a unique highly symmetric
ground-state where each site shares a singlet with any neighbor
connected to it.}: their
ground-states are characterized by regular arrays of singlet
states on
simple bonds or more complicated subsets (4-site plaquettes...).
In the thermodynamic limit the ground-state of a VBC
is degenerate, with a finite degeneracy 
directly related to the symmetry breaking. The VBC
on the $J_1-J_2$ chain (for $\frac {J_2}{J_1} > 0.2411...$)
is doubly degenerate (it breaks the one-step translation symmetry
of the lattice). The simplest VBCs that could be
encountered on a triangular lattice have a 12-fold degeneracy (a
factor  4 comes from the translation symmetry breaking and a
factor 3 from the breaking of $C_3$).
It is a common belief that a VBC has a gap to
all excitations (except indeed at critical points). 

In 2d, the more convincing evidence of such a phase, with a
spontaneous discrete symmetry breaking has been seen 
on the $J_1-J_2$ model on the square lattice 
(see \cite{sz92,kos99} and refs. therein). 
In that case the VBC 
breaks translational symmetry but probably not $C_4$: it is
probably a long range order of $S=0$ plaquettes.

We have a new example of a VBC in the spectrum of
the $J_1-J_2$ model on the honeycomb lattice. When the second
neighbor coupling is strong enough the frustration destroys the
collinear N\'eel long range order and gives birth to a pattern of
valence bonds (with a $Z_3$ symmetry breaking associated to the
three  spatial directions for the dimerized bonds). Our
predictions is based on the scaling of the low lying singlets of
the spectra, it is in agreement with the dimer-dimer correlations
functions measured in the first singlet levels supposed to be
degenerate in the thermodynamic limit\cite{fsl00}.

The nature of
the first excitations of these models (coherent or incoherent magnons)
remains to be elucidated.

\section{Fully gapped Spin-Liquids}

Besides these situations with symmetry breaking and LRO
in a local observable, we have found two other behaviors,
probably generic, with no obvious symmetry breaking, no long range
order in a local order parameter. We call these situations "Spin-Liquids"
and will successively describe both of them.

Let us first describe very quickly the simplest one which has the
following properties 
(an extensive description of the properties of this phase is given
in \cite{mblw98,mlbw99}):

\begin{itemize}
\item A gap to any excitations
\item All spin-spin correlations are short ranged
\item A four-fold degeneracy of the ground-state on a 2-torus
\end{itemize}

We have observed this phase for the first time in the
multiple-spin exchange problem. This effective hamiltonian has
been introduced by Eyring and Thouless to describe the magnetism
of quasi localized fermions (electrons in the Wigner Crystal on
one hand, $^3He$ atoms in solid phase on the other).
The Hamiltonian of the multiple-spin exchange model is given by 
\begin{equation}
H=\sum_{n} (-1)^n J_n (P_n+P_n^{-1})\;\;,\; J_n > 0\;\; , n\ge 2
\label{hamilt}
\end{equation}
where   $J_n$  are   the $n$-spin   exchange  tunneling probabilities
(exchange  coefficients),   $P_n$  and  $P_n^{-1}$  are  the  $n$-spin
permutation operators and their    inverse.  The
alternate  sign in the summation over  $n$ in Eq.~\ref{hamilt} comes
from the  permutation  of   fermions.    In general, the exchange
coefficients decrease with increasing  $n$. The two-spin exchange term
gives the  Heisenberg Hamiltonian up to  a constant: 
\begin{equation}
P_2=2 {\bf S}_i \cdot {\bf S}_j + {1\over 2}
\label{exchange}
\end{equation}
where ${\bf S}_i$ and ${\bf S}_j$ are spins localized  at site $i$ and
$j$,  respectively.  The   three-spin exchange   operator  is equal 
to  a  sum of  two-spin  exchange operators~\cite{mlbw99}.
Thus, the  two and three-spin  exchange terms 
are described (up to a constant) by an Heisenberg hamiltonian
 with an effective coupling constant  $J_2^{\rm eff}=J_2-2J_3$.
Cyclic four-spin exchange gives birth to  new physics: it is 
quartic in spin operators  and has built-in 
frustration. As a result the ground-state of this 
 4-spin exchange hamiltonian on a single plaquette is doubly degenerate with
total spin $S=0$ and $1$! 
At the classical level this hamiltonian displays various
antiferromagnetic phases\cite{k93,km97,msk99}, which are wiped out by
quantum fluctuations, leading in a large range of parameters
 to the Spin-Liquid with the
above-mentioned properties \cite{mblw98,mlbw99}.

The four-fold degeneracy of the ground-state on a 2-torus is a
topological property which was predicted more than ten years
ago\cite{rk88,rc89,rs90}. Contrarily to Oshikawa\cite{o00} we do
not think that it is the signature of long range order in a local
order parameter (see discussion in Misguich\cite{mlbw99}).

Very few things are clearly stated on the excitations of this
Spin-Liquid:
\begin{itemize}
\item Plateaus are expected in the magnetization curves both for
zero and half the saturation magnetization.
\item From the finite size spectra it might be suspected that the
triplet excitations form a "continuum" above the spin-gap and that an
$S=0$ bound-state exists just below the spin-gap.
Up to now, a clear signature of deconfined spinons has not been found
in this Spin-Liquid. Futher investigations appear necessary.
\end{itemize}

As a further hint that might be of importance: short range
correlations are slightly ferromagnetic. In fact this Spin-Liquid phase
appears in the vicinity of a ferromagnetic instability, as can be
seen both in the case of the multiple-spin exchange Hamiltonian on
the triangular lattice, but also for the $J_1-J_2$ model
on the honeycomb lattice\cite{fsl00}.

Experimentally, this phase might be observed in the low density
2d layers of solid $^3He$ where there is 
evidence that the coupling constants might be in the good range
(see \cite{mlbw99,rbbcg97,bbccgrs00,imyf97} and references therein)
and possibly in the Wigner crystal\cite{bc00}.

\section{ Kagom\'e-like Spin-Liquids}

The last situation that we will now discuss is by some details
more puzzling. It has been obtained  by depleting the triangular
lattice,
removing one spin out of four: the lattice which is obtained is called the 
kagom\'e lattice.
Some possible physical realizations are $SrCrGaO$ or various jarosites
~\cite{olitrp88,whmmt98,r94}.
The model has been extensively studied in the classical limit
~\cite{cc91,rcc93,hkb92,chs92a,schb93} 
and it displays an extensive local degeneracy
(i.e. an extensive number of modes of zero energy).
Some authors have pleaded in favor of a selection  of a particular
ground-state by the "order out disorder" mechanism ~\cite{c92}
but the finite size scaling of spin-1/2 spectra contradicts this
assumption ~\cite{lblps97}.
The system is certainly a  Spin-Liquid in the following sense:
\begin{itemize}
\item all the correlations (spin-spin, dimer-dimer, chiral-chiral, ...)
are very small and decay with distance ~\cite{ce92,le93}.
\item the system has a gap for magnetic excitations.
Even if it is not  sure that the largest size (N=36) 
is much larger than all correlation lengths
\footnote{The spin-gap of the $N=21,27,36$ samples seems to be
in a cross-over regime of scaling, 
but the energy per spin still evolves as $N^{-3/2}$!} , 
the finite size scaling studies point to a small  spin-gap:
probably of the order of one tenth of the coupling constant and 
quite likely larger than one twentieth  of the coupling
constant~\cite{lblps97,web98}.

\item the spin-gap is filled by a continuum of singlets and the number of these
singlets  increases exponentially with the system size $N$
~\cite{lblps97,web98} 
as shown in Fig.4.
\end{itemize}

The understanding of the low lying continuum of singlets is still
incomplete but important progresses have been done
recently~\cite{mm00,m98,ze90}. 
Mambrini and Mila ~\cite{mm00} have shown that the 
low lying singlet levels  could be related to a family of short
range dimer coverings of the lattice. It must be emphasized that
short range dimer coverings on this lattice is not necessarily
synonymous with exponential decreasing correlations and gapful
excitations: the recent results of Mambrini and Mila
prove that this is indeed not the case and the comparison between
these results and those of Zeng and Elser~\cite{ze90} point to the
fact that the non-orthogonality of the dimer coverings plays a
major role in this respect. This is an important step forward and
we would like now a simple picture of the first excitations.

Ten years ago, Rokhsar and Kivelson have studied a model of
hard-core dimers on the square lattive~\cite{rk88,lcr96,f91,sv00}.
The phase diagram of
their model displays two VBC phases separated by a critical point.
At the transition point between the two
phases the spectrum of excitations becomes gapless.
The low lying modes are the Goldstone modes associated to a quasi LRO
in the orientation of the dimers.
Let us assume that this hypothesis is  qualitatively  valid on the kagom\'e
lattice and derive its consequences for the number of states below the
spin-gap. Assuming that these Goldstone modes are non-interacting,
it is easy to extract from their dispersion law
($\epsilon(k)=k^n$), the internal energy and entropy versus temperature and thus
the number of states below the first triplet excitation.
The logarithm of this number of states scales as
$N^{\frac{n}{n+2}}$.
In Fig.4  we compare the numerical results with this predicted law
(assuming $n=1$ or $2$). 
Fixing the unknown constants from the measured value for $N=12$,
we see that the prescription is an order of magnitude off at $N=36$:
the $n=1$ mode description  gives $\sim 20$ singlet levels in the spin-gap,
whereas their real number is 
$ 210$~\footnote{ This number is slightly larger than the one quoted in
~\cite{web98} (some degeneracies were incorrectly numbered).}!
With the present pieces of information we must conclude that the description
of the low lying singlets as Goldstone modes associated to a quasi LRO
in the dimer covering is not supported by the numerics
\footnote{A second check has been done on the density of 
states  versus E for $N=36$. Due
to the intrinsic quantum noise of the numerical density of states, it
is impossible  with this last criterium
to really discriminate between the different hypotheses, but the
density of  states implied by a mode description $exp(a E^{\frac {2}{n+2}})$
seems less probable than a pure
exponential or power law ( the $\chi ^2$ criterium is approximately twice as
large).}.
A simple picture of these first excitations is still lacking!

On the other hand, the exponential increase in the number of eigenlevels in a
finite band of energy implies a finite entropy per spin at T=0. The same
phenomena appearing in each spin sector might be at the origin of the anomalous
glassy behavior "without chemical disorder" seen in various compounds
(see \cite{wdvhc00} and refs therein).

Less speculative, the presence of a finite entropy in the singlet channel at low
temperature  is consistent with the observation of 
Ramirez and coworkers\cite{rhw00}
that the low temperature specific heat of $SrCrGaO$ is essentially insensitive
to large magnetic fields (much larger than the
temperature)\cite{smlbpwe00}. This picture is also compatible with the
muons experiments of Uemura {\it et al} \cite{ukkll94} and the elastic spin
diffusion measurement of Lee {\it et al} \cite{lbar96}.

To conclude we must add that these systems seem able to support unconfined
spinons excitations with non zero Chern numbers\cite{web98}.

Up to very recently, we were thinking that this second kind of Spin-Liquid was
associated to special lattices as the kagom\'e or pyrochlore lattices, which at
the classical level display an infinite local degeneracy. Some recent results
bring new pieces of information: we have observed kagom\'e-like spectra on the
triangular lattice when the 3-sublattice N\'eel order is destabilized by a frustrating
4-spin exchange\cite{lmsl00} and we have no evidence of an
infinite local degeneracy at this point.
We now conjecture that the interesting feature common to
these two cases is the appearance of the "kagom\'e-like" phase after destabilization
of a non-collinear long-range order (either due to quantum fluctuations
-when going from the triangular to the kagom\'e lattice by
weakening of bonds\cite{lblps97}-, or to competing frustrating
interactions in the multiple-spin exchange problem\cite{lmsl00}). Contrarily,
destabilization of a collinear order (like in section II) leads to
dimer LRO.

Is this "kagom\'e-like behavior" representative of a true new
phase or of a quantum critical point? This remains an open
question. 
In the two studied cases  we found these characteristics in a rather
wide range of parameters: It is nevertheless impossible to exclude
a finite size effect. If  criticality is to be
viewed as a situation where there is no intrinsic length scale nor
energy scale and a slowing down of all dynamics, it remains unclear
if the proliferation of low lying singlets in this model could not be a
signature of criticality (and the above assumption on
the scaling of the density of states much too naive!).

\section{ Miscellaneous remarks and conclusion}
As a last remark we would like to discuss some questions around
the words and concepts of order and disorder.
The established expression of "order by disorder" (or "order out
of disorder") expresses the fact
that quantum fluctuations (as well as thermal fluctuations for which it
was primarily introduced by Villain) stabilize the simpler and the more
symmetric ordered state amongst the allowed $T=0$ classical
ground-states. But more generally the effect of
long wave-length quantum fluctuations on a (semi-)classical
 ground-state might lead  to still simpler and more symmetric
 situations
which do not break $SU(2)$ (section III) and even any space
symmetries (sections IV and V). In such an issue
the word "quantum disorder" has been used (by opposition to N\'eel
long range order)
 but indeed it should be used with great caution:
the ground-state of the Spin-Liquid of Section IV is
highly symmetric and  very far from a
disordered system.
Only the "kagom\'e-like" state might be qualified as 
"quantum disordered"
(section V). To measure this delicate issue one may look at the
$T=0$ entropy per spin of the ground-state. In the thermodynamic limit,
this quantity scales as ${\mathcal O}\left ( p Ln(N)/N \right
) $ for a N\'eel ordered
state with $p $ sublattices, as $ {\mathcal O}(1/N)$ in the
Valence Bond Crystals (Sec. III) and  in the  fully gapped Spin-Liquid
(Sec. IV),  and  it is  $ {\mathcal O}(1)$  in the "kagom\'e-like"
Spin-Liquids.

In conclusion exact diagonalizations have allowed a
characterization of 4 different and (probably ) generic
ground-states of quantum antiferromagnets in 2d. The
two kinds of Spin-Liquids conjectured by P.W. Anderson in
1973\cite{a73} have been found (the second with the continuum of
singlets seems nevertheless rather different
from the initial conjecture). Possible physical realizations of
these Spin-Liquids exist. There remain many questions on the first
excitations of these Spin-Liquids. Some progress on this subject
can probably be done in computing dynamical structure functions
with exact diagonalizations schemes, but it would also be useful to
devise new analytical approximate tools to look at these
questions:
exact results on the ground-state and low lying levels of the
spectra may hopefully help in achieving
this goal!


\begin{thebibliography}{10}

\bibitem{m91}
E. Manousakis, Rev. Mod. Phys. {\bf 63},  1  (1991).

\bibitem{sz92}
H. Schultz and T. Ziman, Europhys. Lett. {\bf 8},  355  (1992).

\bibitem{blp92}
B. Bernu, C. Lhuillier, and L. Pierre, Phys. Rev. Lett. {\bf 69},  2590
  (1992).

\bibitem{cts99}
L. Capriotti, A. Trumper, and S. Sorella, Phys. Rev. Lett. {\bf 82},  3899
  (1999).

\bibitem{rry89}
J.~D. Reger, J.~A. Riera, and A.~P. Young, J. Phys. Cond. Matt. {\bf 1},  1855
  (1989).

\bibitem{fsl00}
J.-B. Fouet, P. Sindzingre, and C. Lhuillier, in preparation (unpublished).

\bibitem{nz89}
H. Neuberger and T. Ziman, Phys. Rev. B {\bf 39},  2608  (1989).

\bibitem{f89}
D. Fisher, Phys. Rev. B {\bf 39},  11783  (1989).

\bibitem{bllp94}
B. Bernu, P. Lecheminant, C. Lhuillier, and L. Pierre, Phys. Rev. B {\bf 50},
  10048  (1994).

\bibitem{lblp95a}
P. Lecheminant, B. Bernu, C. Lhuillier, and L. Pierre, Phys. Rev. B {\bf 52},
  9162  (1995).

\bibitem{adm93}
P. Azaria, B. Delamotte, and D. Mouhanna, Phys. Rev. Lett. {\bf 70},  2483
  (1993).

\bibitem{lblp95}
P. Lecheminant, B. Bernu, C. Lhuillier, and L. Pierre, Phys. Rev. B {\bf 52},
  6647  (1995).

\bibitem{lblps97}
P. Lecheminant {\it et~al.}, Phys. Rev. B {\bf 56},  2521  (1997).

\bibitem{lmsl00}
W. LiMing, G. Misguich, P. Sindzingre, and C. Lhuillier, Phys. Rev. B {\bf 62},
  6372  (2000).

\bibitem{aklt87}
I. Affleck, T. Kennedy, E. Lieb, and H. Tasaki, Phys. Rev. Lett. {\bf 59},  799
   (1987).

\bibitem{aklt88}
I. Affleck, T. Kennedy, E.~H. Lieb, and H. Tasaki, Commun. Math. Phys. {\bf
  115},  477  (1988).


\bibitem{kos99}
V.~N. Kotov, J. Oitmaa, O. Sushkov, and Z. Weihong, Phys. Rev. B {\bf 60},
  14613,14616  (1999);
L. Capriotti and S. Sorella, Phys. Rev. Lett. {\bf 84},  3173 (2000);
O. P. Sushkov, J. Oitmaa and Z. Weihong, condmat/0007329;
M. S. L. du Croo de Jongh, J. M. J. van Leeuwen and W. van Saarloos,
condmat/0002116. 
This discussion was opened by the  large $N$  study: 
N. Read and S. Sachdev, Phys. Rev. Lett. {\bf 66}, 1773 (1991);
S. Sachdev and N. Read, Phys. Rev. Lett. {\bf 77}, 4800 (1996).

\bibitem{mblw98}
G. Misguich, B. Bernu, C. Lhuillier, and C. Waldtmann, Phys. Rev. Lett. {\bf
  81},  1098  (1998).

\bibitem{mlbw99}
G. Misguich, C. Lhuillier, B. Bernu, and C. Waldtmann, Phys. Rev. B {\bf 60},
  1064  (1999).

\bibitem{k93}
S.~E. Korshunov, Phys. Rev. B {\bf 47},  6165  (1993).

\bibitem{km97}
K. Kubo and T. Momoi, Z. Phys. B. Condensed Matter {\bf 103},  485,489  (1997).

\bibitem{msk99}
T. Momoi, H. Sakamoto, and K. Kubo, Phys. Rev. B {\bf 59},  9491  (1999).

\bibitem{rk88}
D. Rokhsar and S. Kivelson, Phys. Rev. Lett. {\bf 61},  2376  (1988).

\bibitem{rc89}
N. Read and B. Chakraborty, Phys. Rev. B {\bf 40},  7133  (1989).

\bibitem{rs90}
N. Read and S. Sachdev, Phys. Rev. B {\bf 42},  4568  (1990);
N. Read and S. Sachdev, Phys. Rev. Lett. {\bf 66}, 1773 (1991).

\bibitem{o00}
M. Oshikawa, Phys. Rev. Lett. {\bf 84},  1535  (2000).

\bibitem{rbbcg97}
M. Roger {\it et~al.}, Phys. Rev. Lett. {\bf 80},  1308  (1998).

\bibitem{bbccgrs00}
C. Bauerle {\it et~al.}, Physica B {\bf 280},  95  (2000).

\bibitem{imyf97}
K. Ishida, M. Morishita, K. Yawata, and H. Fukuyama, Phys. Rev. Lett. {\bf 79},
   3451  (1997).

\bibitem{bc00}
B. Bernu and D. Ceperley, private com (unpublished).

\bibitem{olitrp88}
X. Obradors {\it et~al.}, Solid State Commun. {\bf 65},  189  (1988).

\bibitem{whmmt98}
A.~S. Wills {\it et~al.}, Europhys. Lett {\bf 42},  325  (1998).

\bibitem{r94}
A.~P. Ramirez, Annu. Rev. Mater. Sci {\bf 24},  453  (1994), and refrences
  therein.

\bibitem{cc91}
P. Chandra and P. Coleman, Phys. Rev. Lett. {\bf 66},  100  (1991).

\bibitem{rcc93}
I. Richtey, P. Chandra, and P. Coleman, Phys. Rev. B {\bf 47},  15342  (1993).

\bibitem{hkb92}
A. Harris, C. Kallin, and A. Berlinsky, Phys. Rev. B {\bf 45},  2899  (1992).

\bibitem{chs92a}
J. Chalker, P. Holdsworth, and E. Shender, Phys. Rev. Lett. {\bf 68},  855
  (1992).

\bibitem{schb93}
E. Shender, V. Cherepanov, P. Holdsworth, and A. Berlinsky, Phys. Rev. Lett.
  {\bf 70},  3812  (1993).

\bibitem{c92}
A. Chubukov, Phys. Rev. Lett. {\bf 69},  832  (1992).

\bibitem{ce92}
J. Chalker and J. Eastmond, Phys. Rev. B {\bf 46},  14201  (1992).

\bibitem{le93}
P. Leung and V. Elser, Phys. Rev. B {\bf 47},  5459  (1993).

\bibitem{web98}
C. Waldtmann {\it et~al.}, Eur. Phys. J. B {\bf 2},  501  (1998).

\bibitem{mm00}
M. Mambrini and F. Mila, cond-mat/0003080 (unpublished).

\bibitem{m98}
F. Mila, Phys. Rev. Lett. {\bf 81},  2356  (1998).

\bibitem{ze90}
C. Zeng and V. Elser, Phys. Rev. B {\bf 42},  8436  (1990).

\bibitem{lcr96}
P. Leung, K. Chiu, and K. Runge, Phys. Rev. B {\bf 54},  12938  (1996).

\bibitem{f91}
E. Fradkin,  in {\em Field Theories of Condensed Matter Systems}, {\em
  Frontiers in Physics}, edited by D. Pines (Addison-Wesley, USA, 1991).

\bibitem{sv00}
S. Sachdev and M. Vojta, J. Phys. Soc. Jpn.  Supp.{\bf  B},  1  (2000).

\bibitem{wdvhc00}
A. Wills {\it et~al.}, cond-mat/0001344; to appear in Phys. Rev.  B., Rap.
Com. {\bf 62},(2000).

\bibitem{rhw00}
A.~P. Ramirez, B. Hessen, and M. Winkelmann, Phys. Rev. Lett. {\bf 84},  2957
  (2000).

\bibitem{smlbpwe00}
P. Sindzingre {\it et~al.}, Phys. Rev. Lett. {\bf 84},  2953  (2000).

\bibitem{ukkll94}
Y. Uemura {\it et~al.}, Phys. Rev. Lett. {\bf 73},  3306  (1994).

\bibitem{lbar96}
S.-H. Lee {\it et~al.}, Europhys. Lett {\bf 35},  127  (1996).

\bibitem{a73}
P. Anderson, Mater. Res. Bull. {\bf 8},  153  (1973).

\end{thebibliography}

\begin{figure}
\centerline{\psfig{figure=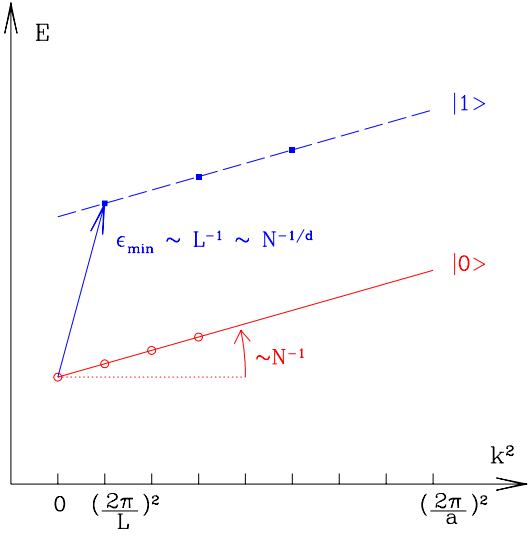,width=7.5cm,angle=0}}
    \caption[99]{ Typical spectrum of a finite size solid.
The first tower of eigenlevels $\left|0\right>$ is associated with c.o.m. motion.
The second tower $\left |1\right>$ describes the softest phonon. 
 }
    \label{Fig1}
\end{figure}
\begin{figure}
\centerline{\psfig{figure=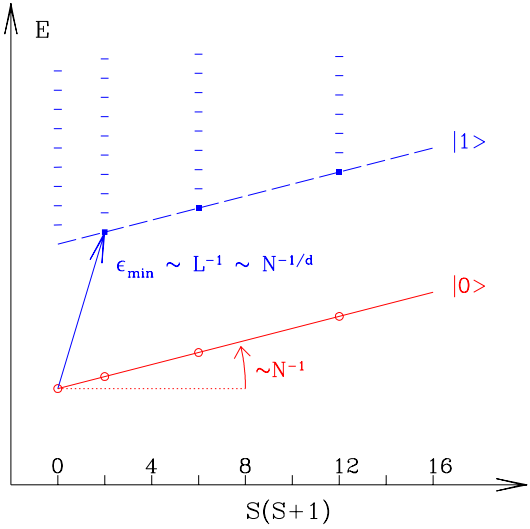,width=7.5cm,angle=0}}
    \caption[99]{ Typical spectrum of a finite size 
collinear antiferromagnet with a N\'eel order.
The tower of eigenlevels $\left|0\right>$ are the QDJS associated with 
the free dynamics of the order parameter.
The second tower  $\left |1\right>$ is associated with the lowest magnon.
 }
    \label{Fig2}
\end{figure}

\begin{figure}
\centerline{\psfig{figure=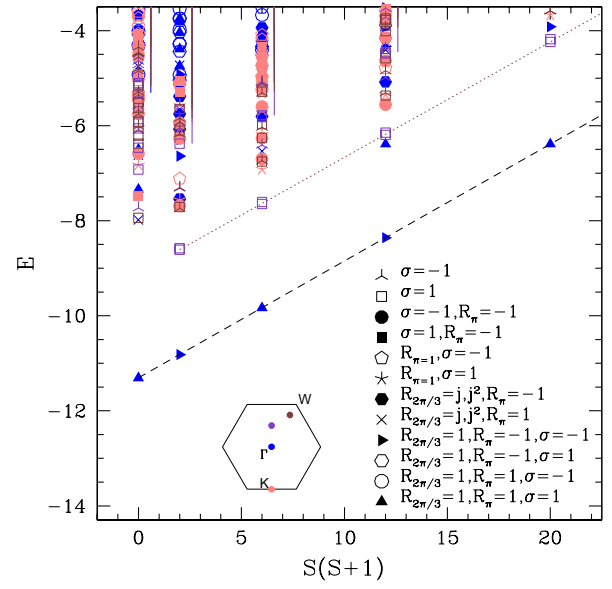,width=8.0cm,angle=0}}
    \caption[99]{ Antiferromagnetic Heisenberg model
on the honeycomb lattice: eigenenergies vs the eigenvalues of ${\bf S}^2$.
The dashed-line is a guide to the eyes for the QDJS. The dotted
line joins the states associated to the first magnon.
There is one QDJS for each $S$
(as expected for a collinear antiferromagnet):
they are ${\bf k=0}$ states,
invariant under a $2\pi/3$ rotation
around an hexagon center, even (odd) under inversion,
odd (even) under a reflexion with respect to an axis going through
nearest neighbor hexagon centers for $S$ even (odd).
 }
    \label{Fig3}
\end{figure}

\begin{figure}
\centerline{\psfig{figure=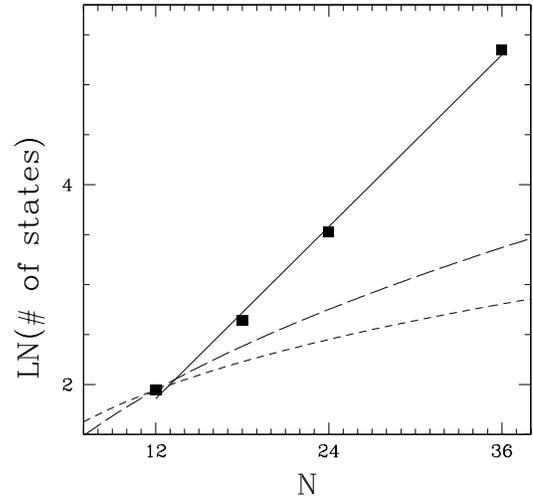,width=7.5cm,angle=0}}
    \caption[99]{ Logarithm of the number of states in the magnetic gap
on the kagom\'e lattice (squares fitted by a continous line) 
vs the size $N$ of the system. 
The observed behavior differs from the
scaling law $exp(bN^{n/(n+2)})$ deduced from a single mode description of the
continuum  with a dispersion law
$\epsilon(k)=k^n$ for $n=1$ (short-dashed line) and $n=2$ (long-dashed line).
 }
    \label{Fig4}
\end{figure} 

\end{document}